\documentclass[aps,pre,onecolumn,showpacs,a4paper]{revtex4}
\input epsf 
\usepackage{graphicx} 
\usepackage{amsmath}
\usepackage{amssymb}
\usepackage{amscd}
\usepackage{soul}
\usepackage{cancel}
\usepackage{color}
\newcommand{\thickone}{\mbox{$1\!\!1$}}

\begin{document}

 \title{Quantum Fluctuation Relations for the Lindblad Master Equation}
\author{R. Chetrite}
 \affiliation{Laboratoire J. A. Dieudonn\'e, UMR CNRS 6621,
 Universit\'e de Nice Sophia-Antipolis, Parc Valrose, 06108 Nice
 Cedex 02, France}
\author{K. Mallick}
 \affiliation{Institut de Physique Th\'eorique, CEA Saclay, 91191
  Gif-sur-Yvette, France}
\date{\today}
 
\begin{abstract}  
      An  open quantum  system interacting  with its environment can be modeled
 under suitable assumptions as a Markov  process, described by a
 Lindblad  master equation. In this work,  we derive a  general  
 set of fluctuation relations for  systems governed  by  a  Lindblad equation.
 These identities provide  quantum versions  
  of Jarzynski-Hatano-Sasa  and Crooks relations.
 In the linear response regime,  these  fluctuation relations yield a  
 fluctuation-dissipation theorem  (FDT) valid for a stationary state
 arbitrarily far from equilibrium.
 For a closed system, this FDT reduces to 
 the  celebrated  Callen-Welton-Kubo formula. 
\end{abstract}

 \maketitle 

\section{Introduction}

           Fluctuations in non-equilibrium systems have been shown to
  satisfy various remarkable  relations 
  \cite{EvansCohen,GallavottiCohen,JarzPRL,JarzPRE,Crooks99,Crooks999,Kurchan2}
   discovered 
  during  the last twenty years.
  These results  have lead to fierce  discussions  concerning
 the nature of heat, work and entropy, raising 
 the  fundamental issue  of
  understanding  the interactions between a given system and its environment
 (e.g., a thermal bath). In the  classical realm, these  problems 
  have been progressively clarified whereas  they are still under
  investigation in the quantum world.

  
Historically,  quantum fluctuation relations were first studied 
by Callen and Welton in 1950 \cite{Callen}.  These authors  derived  a
Fluctuation-Dissipation Theorem for a closed quantum system isolated
from its environment, initially in thermal equilibrium and suddenly
perturbed by a small time-dependent term added to the time-independent
Hamiltonian $H$. This approach was further systematized by Kubo
\cite{Kubo0} in the linear response theory.

 Since 2000, three main
 directions  have emerged for constructing quantum fluctuation relations
which extend the linear response theory of \cite{Kubo0,Callen} and
which are quantum analogues to fluctuation relations for classical
systems. In the following, we rapidly review these three  different routes by emphasizing
 their goals and differences. We also cite useful articles.

  1.  In the first approach, initiated  in 2000 by Yukawa \cite{Yukawa} and
  continued by Allahverdyan and Nieuwenhuizen \cite{Allahverdyan}, a definition for the  quantum 
  work operator is introduced but this operator  does not obey any  fluctuation
  relation.   More precisely, for closed but non-autonomous systems
  (with time-dependent Hamiltonian $H_{t}$), it is proved in
  \cite{Allahverdyan} that  the naive (inclusive) work operator
  $H_{t}^{H}(t)-H_{0}^{H}(0)$ (where the exponent $H$ stands for
  Heisenberg picture) does not satisfy  any fluctuation relation, unless
  $[H_{0},H_{t}]=0$.   This implies the existence of  quantum corrections to the
  Jarzynski relation \cite{Chernyak}. The works of   Bochkov and Kuzovlev
  in 1977 \cite{Boc1}, and of  Stratonovich in 1994 \cite{Str1} can  be viewed as earlier
  attempts in  this direction.

 2.   The second  line of attack  was opened by Kurchan and Tasaki in 2000
\cite{Kurchan,Tasaki}, and  was continued  by many groups
\cite{SMukamel0,Monnai,Talkner1,Talkner2,Gaspard1,
Campisi,Talkner3,JarzynskiQM}. This approach  is based on a different definition
for  the work. The idea is  to  introduce initial {\it and}  final measurements of
the system's energy, according to the quantum-mechanical measurement
postulates. The work (which is viewed as an energy difference) is then a two-point
functional obtained by calculating the difference between the results of the  two
measurements.  This definition of the work differs fundamentally from  the previous one 
(see  \cite{Dere,Engel} for a comparison) and does satisfy quantum  fluctuation  relations. 
The results of this approach  are thoroughly  reviewed in 
\cite{SMukamelReview,CampisiReview}. In summary, we can say that 
 this  definition of the work as a  two-point measurement  has been  applied
in  the following  different contexts: 


 (i)  Closed but non-autonomous systems
prepared in a Gibbs state and isolated from their environment during
their evolution (which is thus unitary).  Kurchan \cite{Kurchan}
studied the time cyclic case with  $H_{t}=H_{T-t}$ ($T$ being  the
period) and proved Jarzynski and Crooks relations in this set-up. 
Tasaki  \cite{Tasaki} generalized this result to the non-cyclic case. Many groups
\cite{SMukamel0,Monnai,Talkner1,Talkner2,Gaspard1} simplified the
theory.  In  particular, in 2007,  Talkner et al. \cite{Talkner1}
clarified   the fact that  work, being  characterized
 through a process with two
measurements,  `is not  an observable' and  
cannot  be represented by any  hermitian operator. 
 The link and the  difference with the older 
Bochkov-Kuzovlev approach \cite{Boc1} (inclusive versus exclusive work)
 was explained by Campisi et al.  in \cite{Campisi2} and the possibility
to perturb the unitary evolution by N-points measurements  was studied by
Callens et al.  in \cite{Callens}. Finally, a  more general relation for a 
`quantum generating functional' was derived by  Andrieux et al. 
 (see equation  12 of \cite{Gaspard1}).

 (ii) The second set-up  corresponds to the 
 general case  of an  open system
 continuously interacting  with its surroundings. This case can be
 formally reduced to the previous situation   considering  the
 system together with  its environment  to be  a  global, closed, 
 system  \cite{Kurchan,Maes,Roeck,Monnai,
 CrooksQM,CrooksQM2, Campisi,Talkner3}.  The main physical  advantage of this
  approach is that one can use the previous  expression of  work  as a  2-points measurements.
   But,  such a  'holistic' 
  approach,    involving   both the system and   its surroundings,  leads  to  fluctuation relations
   that  are  difficult to  assess    experimentally. Indeed, 
  the  environment is  usually large and hardly controllable and only the degrees of freedom
 are experimentally accessible.

(iii) Finally,  the study
of heat-matter exchange for two systems in contact 
\cite{JarzynskiQM,Gaspard2,Saito1,Saito2,SMukamelReview,Saleur}  can be viewed  as a special 
 instance of the previous set-up.  Contact between reservoirs
 at different temperatures and chemical potentials
 lead to  transport of energy and matter.  A famous example is electron counting statistics
\cite{Klich} in which  small nanoscale electronic devices  exchange
electrons.


 3. In a third category of works,  the system is modeled  by an  effective
master equation at the mesoscopic scale. This  approach was pursued in  relatively few
articles  \cite{Yukawa,Maes,SMukamel1,Roeck,CrooksQM}.  The idea is to consider
an  open system  continuously interacting
with  its environment and to  project out the 
 degrees of freedom of the bath  to derive an effective  dynamics for
the system.  Then, it is assumed that the reduced dynamics is described by a
closed evolution equation for the density matrix of the system.  Under
some  further assumptions \cite{Lindblad,Gardiner}, this master
equation can be  brought into a  Markovian form  known as the
Lindblad equation. In the papers  \cite{SMukamel1,Roeck}, the  quantum
master equation is treated  as an effective classical master equation associated
to a pure-jump process;   this  allows the authors to use  the concept of pathwise
trajectory and the trajectorial definition of entropy production. The
 classical  fluctuation relations can then be applied.  This
approach   \cite{SMukamel1,Roeck}, albeit  very powerful because it rests on the highly
developed  field of classical fluctuation relations,
doesn't provide  us with  any  explicit relation involving  quantum
observables. Quantities such  as entropy or work are  defined
along effective classical trajectories 
and   their transposition for the
original quantum system are not at all   obvious (and in fact,  the  effective classical
process is not uniquely defined). 
 A different  philosophy, adopted in \cite{Yukawa,Maes,CrooksQM},  is
to work directly with  the quantum master equation.  However, in  \cite{Yukawa,CrooksQM} 
 the time evolution was discretized 
in an   ad-hoc    manner, and  in  \cite{Maes}  the transition rates  were  given in 
 an arbitrary  way.
   At a conceptual level, one  could object   the relevance of the  approach 3, valid only 
  at a  mesoscopic scale.  Since   fluctuation relations can  be  established 
 at  the microscopic scale (by  the approach 2),   the  theory at  
 the  mesoscopic scale  should simply result  from the proved  microscopic  relations.
  This objection is not valid for two reasons.
  First, one should recall that even   in the classical case, 
 the fluctuation relations were 
 experimentally tested  for  effective stochastic models, valid only  at a  mesoscopic scale
 \cite{Douarche,Liphardt}. A second argument, given 
  by De Roeck in \cite{Roeck},  emphasizes the fact that for a  mesoscopic   quantum  system 
 entropy production is not well defined (in contrast to  the case of a classical system)
  mesoscopic fluctuation relation can not be obtained    by a  coarse-graining procedure.


\hfill\break

  The  present work follows the third approach to quantum fluctuation relations.   We  study the
  non-equilibrium fluctuations  of  an  effective open quantum system
  modeled by a Lindblad  master equation. The Lindbladian evolution is
  a non-unitary dynamics for the density matrix  $\rho_t$ of the
  system, described by a  differential equation  with a generator
  $L_t$ (semi-group property).  This  effective Markovian description
  is  widely used in Quantum Optics \cite{Gardiner}.  But unlike
  \cite{Yukawa,Maes,CrooksQM,SMukamel1,Roeck}, our
  goal, here, is to work directly with  the continuous time Lindblad
  equation,  to define an associated time-reversed dynamics and to
  derive  fluctuation relations with  quantum observables. Therefore the  fluctuations
  relations  we obtain  stem from  structural and symmetry properties of 
   the  Lindblad  master equation.  The key
  results of the present work, given in
  Eqs.~(\ref{fluctuat},\ref{QuantFT},\ref{QuantJarz},\ref{TFDG}),
  represent  an original contribution  to  quantum   non-equilibrium
  statistical mechanics.

  
 Our strategy will be to use  a  suitable deformation
 of the master equation,   which will allow us to  prove  a generic relation 
 amongst correlation functions,  a kind
  of book-keeping formula  which  is   a  quantum analog of Jarzynski-Hatano-Sasa
 relation and Crooks relation.  
  Furthermore,  by  a  lowest order expansion, 
 we derive a  generalized fluctuation-dissipation theorem 
 valid in the vicinity of  a quantum  non-equilibrium steady state.  
 For the special case of a closed system, our approach retrieves
 previously known work identities \cite{Kurchan,Tasaki,Talkner1,Gaspard1,Yukawa,Maes} as well as  
 the quantum equilibrium  fluctuation-dissipation theorem \cite{Callen,Kubo}.

   The outline of this work is as follows. In section~\ref{Sec:Master}, we recall some basic
 properties of the Lindblad equation which represents an effective Markov evolution for
 a system in contact with an environment. We write a formal solution for the  Lindblad equation
 and give the expression of multi-time correlation functions. In section~\ref{sec:QJarzynski}, we 
 prove the Quantum Jarzynski-Hatano-Sasa relation associated
  with the Lindblad equation. We use this relation
 to derive a steady-state quantum Fluctuation-Dissipation
  theorem that generalizes the Kubo-Callen-Welton
  formula (which applies only  to  a closed system
  in the vicinity of thermal equilibrium). In section~\ref{sec: Tasaki-Crooks},
 we investigate the properties of a given   Lindblad evolution under time-reversal.
 This allows us to prove for open quantum systems
 a general version of the Tasaki-Crooks relation. Concluding remarks are
 given in   section~\ref{sec:Conclusion}. Technical details are deferred to the appendices.

\section{Master equation for Quantum Markov Dynamics}
\label{Sec:Master}

   Consider a quantum system $S$  in contact with a thermal reservoir (or environment) $R$. The total
   system  $S + R$ is closed. Its evolution is unitary and the total density matrix
   evolves according to the quantum  Liouville equation governed by the total  Hamiltonian
 $H_{{\rm total}}$ which can be broken into three pieces: the Hamiltonian
 of the (small) system $H_S$, the  Hamiltonian of the reservoir $H_R$ and the
 interaction Hamiltonian between the system and its environment $H_I$. We are 
  interested in the degrees of freedom of  $S$ and therefore  we  would
 like  to have at our disposal an evolution equation for the density matrix $\rho_t$  of the 
 quantum system $S$ alone,  the degrees of freedom of the reservoir being  traced out.
  Generically, such an equation is non-local in time: the coupling $H_I$ induces  memory effects.
 However, under some specific assumptions, a differential equation of first order
 with respect to time can be derived for  $\rho_t$: one must assume that 
 (i)  the  full system $S + R$
 is prepared in a correlation free state; (ii) the reservoir  $R$ is large enough so that
 it has a very short memory time $\tau_c$ (more precisely one must have 
$\tau_c \ll  \hbar/{|H_I|} $ where $|H_I|$ is the typical order of magnitude of the interaction
 $H_I$ matrix elements \cite{Haroche}). This second assumption is the crucial Markov hypothesis:
 when it is fulfilled a memoryless and coarse-grained description of the system $S$ becomes
 possible.  This condition is  generally  satisfied in  Quantum Optics \cite{CCT,Haroche}. 

  There are  two main methods to derive the master equation for 
  quantum Markovian dynamics. 
 One way is to make  a precise model for the reservoir (typically an infinite  set of
 quantum oscillators) and  to  eliminate  explicitly   the 
environmental degrees of freedom. The Markov approximation can be analyzed and justified
 precisely \cite{CCT}. Another possibility is to study the structural properties of the 
 'quantum map'  that carries  the density matrix $\rho_t$ at time $t$   into 
 the density matrix  $\rho_{t +dt}$ at time $t + dt$.  Such a  map must be linear, 
  hermiticity preserving,  
  trace conserving  and  positive. In fact, this  map  lifted to an operator
  on the total system $S + R$ (for any given  environment)  must remain 
  positive (this stronger  requirement  is called complete positivity)  \cite{Haroche, Gardiner}.
 These  physically reasonable  conditions are stringent enough 
to mathematically constrain  the possible
 forms  of the evolution equation  of a quantum Markov 
dynamical system \cite{Lindblad}. The resulting
 equation is called a {\it Lindblad  equation}. Its generic form,  equation~(\ref{eq:master0}),
 and some of its basic properties are discussed below.

\subsection{Some properties of the Lindblad equation}

    We  consider a quantum system prepared initially  with a  density matrix  $\pi_0$.
    Because of its interactions with its environment, the  density matrix of the system,
   becomes a function of  time, and will be denoted by $\rho_t$. In the present work, we 
 adopt the framework of Quantum Markovian Dynamics.
 The evolution of  $\rho_t$ is thus given  by  a  Lindblad master  equation, which can be written
 in  the  generic form \cite{Haroche, Gardiner}: 
 \begin{equation}
 \partial _{t}\rho_{t}  
 = - i[H_t,\rho _{t}] + \sum_{i=1}^{I} \left( V_{i} \rho _{t} V_{i}^{\dagger }
  - \frac{1}{2} V_{i}^{\dagger } V_{i} \rho _{t} 
  - \frac{1}{2}   \rho _{t} V_{i}^{\dagger}V_{i} \right)  \,.
\label{eq:master0}
 \end{equation} 
  On  the right-hand side of this
  equation, the  first term $-i[H_t,X]$ is the conservative part 
  where $H_{t}$ is  the Hamiltonian  of the system that may depend on time.
  The other terms  represent  the interactions of the system 
 with its  environment (also called the `bath') and also  represent the effect
 of  measurements  (i.e.  dissipation and  (de)coherence effects).
 The operators  $V_{i}$ are called
 the {\it Kraus operators}, they are not necessarily 
 hermitians  and   may  depend explicitly  on time. The  Kraus 
 number $I$ depends on  the system considered. If the system
 under consideration is closed, all  Kraus operators  vanish  identically,
  $V_{i} \equiv 0$,  and the Lindblad master equation reduces to the
 quantum version of the Liouville equation. Equation~(\ref{eq:master0})
can be written symbolically as 
\begin{equation}
 \partial _{t}\rho _{t} = L_{t}^{\dagger}  \rho _{t} \, , 
 \label{eq:master1}
 \end{equation} 
  where we have introduced the  Lindbladian superoperator
 $L_{t}^{\dagger}$ which acts on the density matrix  $\rho_t$ and generates
 its time-dynamics. We emphasize that  $L_{t}^{\dagger}$ is 
 a superoperator because it is  a linear map in the space of operators. 
 The fact that we have used   in equation~(\ref{eq:master1})
 the symbol  $L_{t}^{\dagger}$ for  the Lindbladian  rather than the more usual notation
 $L_{t}$ is purely  a matter of convention: this  will allow us to write
 some expressions of time-ordered  correlations of observables
  in a simpler manner (because  $L_{t}^{\dagger}$ acts on the 
 density matrix  $\rho_t$  and  its conjugate superoperator  $L_{t}$
  acts on observables).  More precisely, the 
  space of operators  is  endowed  with  the following Hilbert-Schmidt  scalar product 
 $(Y,X) = Tr(Y^{\dagger} X)$, where  $X$ and $Y$ are
   arbitrary  operators  and $Y^{\dagger}$ is  the hermitian conjugate of $Y$.
 This allows us to define  a pair of  adjoint  superoperators
$L_{t}$  and  $L_{t}^{\dagger}$ as follows
\begin{equation}
 (Y, L_{t} X) =  Tr(Y^{\dagger} ( L_{t} X)) =  (L_{t}^{\dagger} Y, X)
 =  Tr( (L_{t}^{\dagger} Y)^{\dagger} X) \, . 
\end{equation} 
 A simple calculation allows us to write
  \begin{equation}
 L_t  X = i[H_t,X] + \sum_{i=1}^{I}
\left( V_{i}^{\dagger }XV_{i}  - \frac{1}{2} V_{i}^{\dagger }V_{i}X  - \frac{1}{2}
 XV_{i}^{\dagger}V_{i} \right)  \,. 
\label{def:Lindbladian}
 \end{equation}   
  Two important properties of the Lindbladian $ L_t$ are $  L_t 1 = 0$ (Trace conservation)
 and $  L_t (X^{\dagger}) = (L_t X)^{\dagger}$.

  The Lindblad  equation is extensively used in Quantum Optics. A simple example is a two-level
 atom emitting a photon in free space. The density matrix $\rho_t$ is  a 2 by 2 matrix
 and the Kraus operators reduce to Pauli lowering and rising operators. The Lindblad
 equation is then simply a set of four coupled first order differential  equations \cite{Haroche}.

\subsection{A formal solution of  the Lindblad equation}

  The quantum Master equation~(\ref{eq:master1})   can be solved
 formally by introducing the evolution  superoperator $P_{0}^{t}$:
\begin{eqnarray}
\rho_{t} =  \left(P_{0}^{t} \right)^{\dagger} \pi_{0} \, ,
\label{eq:FormalSol}
 \end{eqnarray} 
 where $\pi_{0}$   represents the density-matrix at the initial time. 
  The evolution  superoperator $P_{s}^{t}$  between the two times
 $ s \le t$  is  defined by 
\begin{eqnarray}
 P_{s}^{t}  &=& \overrightarrow{\exp }\left( \int_{s}^{t}du\, L_{u}\right)
 =  1 +  \sum_{n=1}^{\infty}\int_{s\leq t_{1}\leq t_{2}\leq\ldots\leq t_{n}\leq t}
 \prod_{i=1}^n dt_{i}\, \, L_{t_{1}} L_{t_{2}}\ldots L_{t_{n}} \, .
\label{eq:ExpOrd1}
 \end{eqnarray} 
 In this time-ordered exponential, time is increasing from left to right.
  This symbolic writing will be very useful
 to perform formal calculations and to write perturbative expansions.
  Let us recall how  equation~(\ref{eq:FormalSol}) is proved.
  First, we  observe
 that this  equation is true at $t=0$ because $P_{0}^{0}$ is the identity
 operator. Then, from the  time-ordered exponential~(\ref{eq:ExpOrd1}),
 we find  
 $\frac{d}{dt}P_{s}^{t} =  P_{s}^{t}  L_{t}$. This leads us finally to
 \begin{eqnarray}
 \frac{d}{dt}\rho_{t} = 
  \left( L_{t}^{\dagger}  (P_{0}^{t})^{\dagger} \right) \pi_{0}
 =  L_{t}^{\dagger} \rho_{t}  \, .
 \end{eqnarray}  
 Thus,   $\rho_{t}$ satisfies the Lindblad equation~(\ref{eq:master1})
 with initial condition $\pi_{0}$.  We note that this technique of proving
 an identity between operators, such as equation~(\ref{eq:FormalSol}),
 by showing that both operators are solutions of the same (first
 order) differential equation with the same initial condition, will
 be used repeatedly in this work. 

\subsection{Expression for multi-time correlations}

  Using the evolution operator, we can write a general expression for 
  multi-time correlations of  different observables.
 For $0\leq t_{1}\leq t_{2}\leq \ldots \leq t_{N}\leq t$, the
  time-ordered   correlation  of observables  $O_{0},O_{1},O_{2}...O_{N}$
is given by
\begin{eqnarray}
&\left\langle O_{1}(t_{1})O_{2}(t_{2}) \ldots O_{N}(t_{N})\right\rangle_{ \pi _{0}}  
&=  Tr\left( \pi_{0} P_{0}^{t_{1}}O_{1}P_{t_{1}}^{t_{2}}O_{2} \ldots 
P_{t_{N-1}}^{t_{N}}O_{N}\right) \, .
\label{def:Correlations}
\end{eqnarray}
 A justification  of this expression can be found in \cite{Gardiner} or in
  \cite{Breuer}:  One starts by  the correlation of two
 operators  at two different times.  This correlation
 can be evaluated in the Heisenberg representation
 by  using the full Hamiltonian of the system {\it plus}  its
 environment. In order to obtain an expression that refers to the                                  
 system alone, the partial trace over the environment has to be performed.
 Using the same assumptions (factorization of the initial condition
 and weak coupling) that lead to  the Lindblad  Master
 equation~(\ref{eq:master1}), one shows that the  time-ordered 
 two-time correlation  function satisfies  an evolution equation which is the 
 dual of  the  equation~(\ref{eq:master1}), i.e. it is governed by the Lindblad
 operator $L_{t}$. This proves  the   formula~(\ref{def:Correlations})
  for  $N=2$. The general case is  then obtained by  induction. 
 Besides, a rigorous proof of  the formula~(\ref{def:Correlations})
 is given  in \cite{Bauer} for a
 toy model of one harmonic oscillator in a bath of independent  harmonic oscillators. 

 We recall that,  in the formula~(\ref{def:Correlations}), the operator
   $\pi_0$
 represents the  initial density matrix of the system 
 and  emphasize that  the superoperator
   $P_{t_{i}}^{t_{i+1}}$ operates on all the terms to its right. For example,
 for $N=3$,  the  explicit expression, with all the required
 brackets,  is given by:
$$\left\langle O_{1}(t_{1})O_{2}(t_{2})O_{3}(t_{3})\right\rangle_{ \pi _{0}}  
 =   Tr\left( \pi_{0}  P_{0}^{t_{1}}(O_{1} P_{t_{1}}^{t_{2}}(O_{2}
P_{t_{2}}^{t_{3}}(O_{3})))\right) \, .$$

\subsection{The accompanying density matrix}

 We suppose that the initial density matrix  $\pi _{0}$ of 
the system  satisfies 
 $L_{0}^{\dagger}\pi _{0} = 0\,.$  For example, 
 the system is prepared in a thermal state at temperature
 $T = \frac{k}{\beta}$, where $k$ is Boltzmann's constant and  
 its  initial density matrix is 
 $\pi _{0}=Z_{0}^{-1}\exp (-\beta H(0)).$  If the
 system is closed and the Hamiltonian is constant in time, then the
 density matrix  does not vary with time:  $\pi _{t}=\pi _{0}$.
 However, if the  Hamiltonian  $H_t$  changes with time and/or  if the interactions
 with the environment are taken into account, the density matrix 
 evolves according to equation~(\ref{eq:master1}) and the system is
 out of equilibrium.  
  In particular,  at time $t$, the system does {\it not} lie in the
  kernel  of the  time-dependent generator 
  $L_{t}^{\dagger}$.  For example,  at time $t$,   a closed system with 
  time-dependent  Hamiltonian  $H_t$, is not in the Gibbs state
  $Z_{t}^{-1}\exp (-\beta H_{t})$. (The same feature is  also true,
  of course, in classical mechanics). 

  Nevertheless, given a  time-dependent  Lindbladian  $L_{t}$,
  it is useful, following   \cite{Hanggi}, to  associate to it 
  the  {\it accompanying}  density-matrix $\pi_t$ that satisfies  
  $L_{t}^{\dagger} \pi_t = 0$. Physically,  $\pi_t$  represents
 the stationary  state  in a  system where time is  frozen
 at  its instantaneous value $t$.   For example, for a closed
 system, we have  $\pi _{t}=Z_{t}^{-1}\exp (-\beta H_{t}).$
 However, as we emphasized  above, 
 at time $t$ the {\it true}  density matrix  of the  system  differs from
 the accompanying  density-matrix: $ \rho_t \neq \pi_t$. The reason is that
$\pi_t$ depends   on time, and therefore
   it can  not satisfy  Eq.~(\ref{eq:master1}): 
 $$\frac{d}{dt}\pi_t \neq  0  \,\,\,
 \hbox{ whereas }  \,\,\,  L_{t}^{\dagger} \pi_t = 0 \, . $$

\section{A  Quantum Jarzynski-Hatano-Sasa Identity for  Lindblad dynamics}
\label{sec:QJarzynski}

 A  key idea that lies at the heart  of the   Jarzynski identity 
 in the classical case  is to consider  non-equilibrium 
 averages over weighted trajectories. This
 crucial feature was  clearly recognized and   stated in the 
 very early works \cite{JarzPRL, JarzPRE}.
 If the weighting factor is chosen
 to be the exponential of the 
 work performed on the system, then   weighted averages
 along a  non-equilibrium  process between times  0 and  $t$ can be  reduced  to 
 thermal  averages performed with the accompanying Gibbs measure at time $t$.
   An equivalent formulation  due to  Hummer and Szabo \cite{Hummer}
 is to  consider  an auxiliary   system governed by 
 {\it a fictitious   dynamics}, constructed  in such  a  way
 that  at each time $t$  the  auxiliary system  lies in 
 the accompanying steady-state  measure at time $t$ of the
 initial system.   Thus,  averages on the fictitious
    system  can be written as   steady-state averages in  
  the initial system. Besides,  using the Feynman-Kac formula, 
 Hummer and Szabo  showed   that averages
 over  the fictitious  system are  given by  averages over 
  the initial system
 weighted by the Jarzynski factor (the exponential of the work).  Eliminating
 the auxiliary system between the two equalities leads to the  Jarzynski 
 identity.

 It is important to note that exactly the
  same idea of considering a   modified  dynamics  appears in the proof
 by Kurchan of the  Gallavotti-Cohen relation for Langevin dynamics 
 \cite{Kurchan2} and  also in the general derivation of the 
 Gallavotti-Cohen symmetry for Markovian systems by Lebowitz and Spohn
 \cite{LeboSpohn}.

 Our aim  in this section is to prove 
 a  Jarzynski-Hatano-Sasa  identity for a quantum Markovian dynamics
  by using a similar  technique:  the Lindblad dynamics is deformed   so that 
 the  accompanying  density-matrix $\pi_t$ of the initial system becomes
 the true density-matrix  at time $t$ of the fictitious  auxiliary  system.
 Then,  an  operatorial version  of the Feynman-Kac formula will allow
 us to relate expectations values over   the auxiliary   system to 
 averages on the initial system, leading us to  a quantum version
 of the Jarzynski-Hatano-Sasa  identity. In the case of  a closed system, we shall show
 that this identity is equivalent   to relations that
 were previously known. Finally, we deduce from our general result
 a fluctuation-dissipation theorem valid for an arbitrary steady state.


\subsection{A modified dynamics for the accompanying density matrix}

 The accompanying density-matrix  $\pi_t$ does not obey  the
 Lindblad equation Eq.~(\ref{eq:master1}).  However, $\pi_t$ is a tautological solution
 of  the modified evolution  equation
   \begin{equation}
  \partial _{t}\pi_t = \left(L_{t}+\pi _{t}^{-1}
  \left(\partial _{t}\pi _{t}\right) \right)^{\dagger} \, \pi_t \,.
 \end{equation}
 We introduce  the non-stationary operator
  \begin{equation}
 W_t = -(\pi_t)^{-1} \left(\partial_t \pi_t\right) \,,
 \label{W}
  \end{equation}
 which reduces in the classical limit to the rate of injected  power,
 and we define a {\it modified}  superoperator as follows
 \begin{equation}
     L_{t,1}=L_{t} + (\pi_t)^{-1} \left(\partial_t \pi_t\right)  =  L_{t}  - W_t \,,
 \label{def:Lind1}
 \end{equation}
 where   $W_t$ acts on a given density-matrix by  
a multiplication  on  the left.
 Then, equation~(\ref{eq:ModifLind}) can be rewritten as
  \begin{equation}
 \partial _{t}\pi_t =  L_{t,1}^{\dagger} \, \pi_t \ . 
 \label{eq:ModifLind}
  \end{equation}
  The   formal solution of this equation  is given by 
 \begin{equation}
 \pi_t =   \left(P_{0,1}^{t}  \right)^{\dagger} \,   \pi_0   \, ,
 \label{eq:ModifFormalSol}
 \end{equation}
 the  modified evolution superoperator $P_{s,1}^{t}$ being  defined as 
\begin{equation}
  P_{s,1}^{t} = \overrightarrow{\exp }\left( \int_{s}^{t}   L_{u,1}\, du
\right) = \overrightarrow{\exp
  }\left( \int_{s}^{t} \left( L_{u} +\pi _{u}^{-1} \left(\partial
  _{u}\pi_u \right) \right) \,du\right) \, .
 \label{eq:ModifPropag}
 \end{equation}
  We observe that equations~(\ref{eq:ModifFormalSol}, \ref{eq:ModifPropag}) for the 
 accompanying matrix  are   similar to equations~(\ref{eq:FormalSol}, \ref{eq:ExpOrd1}) 
for the 'true' density matrix of the system. 

 Now consider  an arbitrary observable $A$. Then,  using equations~(\ref{eq:ModifFormalSol}),
 we can write 
\begin{equation}
Tr\left( \pi_{0}P_{0,1}^{t}A\right)  =
  Tr \left( \pi_{t} A\right) \, .
 \label{eq:AverageModif}
\end{equation}
  This identity  means that averages  for the  fictitious  evolution
 which are  performed by using the  modified  evolution superoperator $P_{s,1}^{t}$ reduce
 to averages  performed with the accompanying density-matrix at time $t$. 

\subsection{Proof of the Quantum Jarzynski-Hatano-Sasa Relation}
\label{subsec:ProofJ}

   A quantum version of the Jarzynski-Hatano-Sasa  Relation will be obtained by 
  relating  the  auxiliary evolution superoperator $P_{s,1}^{t}$   
 to  the initial  evolution  superoperator $P_{s}^{t}$. To  achieve this aim, we need
 to prove an extension of the Feynman-Kac formula.

   We  write a  Dyson-Schwinger series expansion for
  $P_{0,1}^{t}$,  
  considering  $W_t$  to be  a perturbation of the Lindbladian  $L_{t}$:
 \begin{equation}
  P_{0,1}^{t} = \sum_{n}  (-1)^n \int_{0\leq t_{1}\leq t_{2}\leq\ldots\leq t_{n}\leq t}
\, \,  \prod_{i=1}^n dt_{i}\, \, 
    P_{0}^{t_{1}} W_{t_{1}} P_{t_{1}}^{t_{2}} W_{t_{2}}  \ldots 
 P_{t_{N-1}}^{t_{N}}  W_{t_{N}} P_{t_{N}}^{t}  \, ;
\label{DysonSchwinger}
\end{equation}
  we  recall that the superoperator
   $P_{t_{i}}^{t_{i+1}}$ operates on all the terms to its right.
  This well-known  formula \cite{Orland}  can be proved  by showing
  that both sides of the equation 
  satisfy the same differential equation and are identical at $t=0$. 
  Inserting this expansion on  the r.h.s. of equation~(\ref{eq:AverageModif}),
  we obtain 
\begin{eqnarray}
Tr\left( \pi_{0}P_{0,1}^{t}A\right)  =
   \sum_{n}  (-1)^n \int_{0\leq t_{1}\leq t_{2}\leq\ldots\leq t_{n}\leq t}
\, \,  \prod_{i=1}^n dt_{i}\, \, 
  Tr\left( \pi_{0}  P_{0}^{t_{1}} W_{t_{1}} P_{t_{1}}^{t_{2}} W_{t_{2}}  \ldots 
 P_{t_{N-1}}^{t_{N}}  W_{t_{N}} P_{t_{N}}^{t} A \right)  \, .
\end{eqnarray}
 Rewriting the trace term inside the integrals 
  as a multi-time  correlation function via  Eq.~(\ref{def:Correlations}), 
 we find 
\begin{eqnarray}
Tr\left( \pi_{0}P_{0,1}^{t}A\right)  =   
\sum_{n}  (-1)^n \int_{0\leq t_{1}\leq t_{2}\leq\ldots\leq t_{n}\leq t}
\, \,  \prod_{i=1}^n dt_{i}\, \, 
\left\langle  W_{t_{1}}(t_{1}) W_{t_{2}}(t_{2}) \ldots  W_{t_{N}}(t_{N})A(t)\right\rangle_{ \pi _{0}} 
  \, .
\label{eq:beforeRessum}
\end{eqnarray}
 We note that on the r.h.s. of   Eq.~(\ref{def:Correlations}), the operators $O_{i}$  depend
 on the time $t_i$;  this  time 
 dependence is written as an argument  $O_{i}(t_i)$. Here we have  $O_{i} =  W_{t_{i}} $ 
 which  already depends   on time. A  supplementary time 
 dependence is introduced   through the use of 
 Eq.~(\ref{def:Correlations}), which now  appears as  $W_{t_{i}}(t_{i})$.  
 By linearity, equation~(\ref{eq:beforeRessum}) is identical to 
\begin{eqnarray}
Tr\left( \pi_{0}P_{0,1}^{t}A\right)  =   
\left\langle  \left\{ \sum_{n}  (-1)^n \int_{0\leq t_{1}\leq t_{2}\leq\ldots\leq t_{n}\leq t}
\, \,  \prod_{i=1}^n dt_{i}\, \, 
 W_{t_{1}}(t_{1}) W_{t_{2}}(t_{2}) \ldots  W_{t_{N}}(t_{N}) \right\} \,  A(t)\right\rangle_{ \pi _{0}} 
  \, .
\label{eq:beforeRessum2}
\end{eqnarray}
 The term between curly brackets can be resummed as a  time-ordered exponential 
\begin{eqnarray}
Tr\left( \pi_{0}P_{0,1}^{t}A\right)  =   \left\langle 
  \overrightarrow{\exp }
  \left( -\int_{0}^{t} W_{u}(u) \, du  \right) A(t)\right\rangle_{ \pi _{0}}  \, .
\label{eq:afterRessum}
\end{eqnarray}
This formula is an extension of the Feynman-Kac formula for quantum Markov semi-groups.
We emphasize that , due to non-commutativity of operators, the  exponential 
 that appears in the usual  Feynman-Kac formula is replaced here  by a 
 time-ordered exponential. Moreover, we remark that, although 
 there  exist many generalizations of the Feynman-Kac formulae 
\cite{Accardi,DelMoral}, the present one seems to be original.
 Finally, using equation~(\ref{eq:AverageModif}),  the following relation
 is derived:
 \begin{equation}
  Tr \left( \pi_{t}A\right)=
\left\langle \overrightarrow{\exp }
\left( -\int_{0}^{t} W_{u}(u) \, du \right)  A(t) \right\rangle_{ \pi _{0}}  \, . 
\label{Jara}
\end{equation}
 This   identity  is  a quantum extension 
 of the classical Jarzynski-Hatano-Sasa identity
 and is  one of the main  results of the present work.  In particular, 
if  we take $A  = \thickone$  
 then  Eq.~(\ref{Jara}) becomes 
 \begin{equation}
  \left\langle \overrightarrow{\exp }
 \left( -\int_{0}^{t} W_{u}(u) \, du \right) \right\rangle _{ \pi _{0}} = 1  \, , 
\label{QuantJarz}
\end{equation}
 where we use the fact that $  Tr \left( \pi_{t}\right)= 1$.
 Note that if  we interpret the mean values as classical
 averages and the operators as commuting c-numbers,
 then Eq.~(\ref{QuantJarz}) reduces to  the classical Jarzynski-Hatano-Sasa relation.
    The relation~(\ref{QuantJarz}) can be interpreted as  a kind
  of a  book-keeping formula which allows us to write identities amongst
 correlation functions. We emphasize that the operator $W_{t}$ is not Hermitian:
 this is a signature of the fact that 'Work is not an observable' \cite{Talkner1}.
 However, for the special case of a closed system, the  work-term can be written
 as a product of two observables, as will be shown below, allowing us to 
 retrieve   previously known work identities
 \cite{Kurchan,Tasaki,Talkner1,Gaspard1,Yukawa,Maes}.
 Besides, from  a first order expansion of  Eq.~(\ref{QuantJarz})
  we shall  derive 
 a  generalized fluctuation-dissipation theorem  valid
 in the vicinity of  a quantum  non-equilibrium steady state.

\subsection{The  case of a  closed quantum system}

  We consider the special case of
 a closed,  isolated system, governed  by a time-dependent Hamiltonian.
 The Lindbladian reduces to the Liouville
 operator,  $L_t .  X =i[H_t,X],$  and the 
 evolution of the system is unitary.  For a closed system,
 the evolution superoperator  $P_{0}^{t}$ acts   on an observable  
 $X$ as follows
 \begin{equation}
 P_{0}^{t} X =\left( U_{0}^{t}\right)^{\dagger} \,  X \,  U_{0}^{t}  \, ,
 \label{eq:closedSuperP}
 \end{equation}
 where the  unitary evolution operator is defined as
 \begin{equation}
  U_{0}^{t}=\overleftarrow{\exp }\int_{0}^{t}du\, \left( -iH_{u} \right)  \, . 
  \label{eq:closedU}
 \end{equation}
  Here, the  arrow  pointing towards the left  over the
  time-ordered exponential indicates that early times are written on the right
 and   later times  on the left.
   
 The image of  $X$ by 
 the evolution superoperator $P_{0}^{t}$ defines   the Heisenberg
 operator  $X^{{\mathcal H}}(t)$  where the upper-script ${\mathcal H}$ stands
 for Heisenberg:
  \begin{equation}
 P_{0}^{t} X =  X^{{\mathcal H}}(t)  \, .
 \label{eq:closedSuperP2}
 \end{equation}
  We also note that  $ P_{0}^{t}$  is a multiplicative superoperator i.e. 
 for any two observables $X$ and $Y$ we have 
 \begin{equation}
 P_{0}^{t} \left(  X Y \right) =  P_{0}^{t} (X) \,  P_{0}^{t} (Y) \, .
  \label{eq:multipli}
 \end{equation}
 (This  is not true in general for an open system.) Thanks to  this  property,
  the r.h.s of  the general  expression~(\ref{def:Correlations})
 for multi-time correlations can be evaluated and we obtain:
\begin{eqnarray}
&\left\langle O_{1}(t_{1})O_{2}(t_{2}) \ldots O_{N}(t_{N})\right\rangle_{ \pi _{0}} 
&=  Tr\left( \pi_{0} \, O_{1}^{{\mathcal H}}(t_{1})O_{2}^{{\mathcal H}}(t_{2})
  \ldots O_{N}^{{\mathcal H}}(t_{N})\right) \, .
\label{def:ClosedCorrelations}
\end{eqnarray}
If we substitute this expression in equation~(\ref{eq:beforeRessum}) and retrace the steps
from      equation~(\ref{eq:beforeRessum2})      to equation~(\ref{Jara}), we find 
 that for a closed system the  quantum  Jarzynski-Hatano-Sasa  relation  can be written as 
 \begin{equation}
  Tr \left( \pi_{t}A \right)=  Tr\left( \pi_{0}  
 \overrightarrow{\exp }
\left( -\int_{0}^{t} W_{u}(u)^{{\mathcal H}}  \, du \right)  A^{{\mathcal H}}(t) \right) \, , 
\label{ClosedJara}
\end{equation}
 where from equations~(\ref{eq:closedSuperP2})
 and (\ref{eq:multipli}), we have 
\begin{equation}
 W_{u}(u)^{{\mathcal H}} =  -\left(\pi_{u}^{-1}\right)^{{\mathcal H}}(u)
 \left( \partial_{u}\pi_{u}\right)^{{\mathcal H}} (u)  \, .
\label{Def:WHeisen}
\end{equation}
 We also recall that  for a closed system,
   the accompanying density is given by  $ \pi_t = Z_{t}^{-1}\exp (-\beta H_{t})$.
 Equation~(\ref{ClosedJara}) can be simplified thanks 
 to the following operator identity, which is  proved in Appendix~\ref{Proof:IdentiteFermee}:
 \begin{equation}
\overrightarrow{\exp }
\left( -\int_{0}^{t} W_{u}(u)^{{\mathcal H}}  \, du \right) =   (\pi_0 )^{-1}\pi_t^{{\mathcal H}}(t) \, .
 \label{IdentiteFermee}
\end{equation}
   Substituting equation~(\ref{IdentiteFermee}) in equation~(\ref{ClosedJara}), the following (tautological)
 identity is obtained 
\begin{equation}
 Tr \left( \pi_{t}A \right)=  Tr\left\{ \pi_{0} 
  \left(  (\pi_0 )^{-1} \pi_t^{{\mathcal H}}(t)    \right)  A^{{\mathcal H}}(t) \right\}   \, . 
  \label{QJarz2}
\end{equation}
   Taking  $A  = \thickone$,   we end up
 with the quantum Jarzynski relation for closed systems
 as first found  by Kurchan and Tasaki \cite{Kurchan,Tasaki}:
 \begin{eqnarray}
 Tr\left( \pi _{0}\exp (\beta H_{0}^{{\mathcal H}}(0))
 \exp \left( -\beta H_{t}^{{\mathcal H} }(t)\right) \right)
  =\frac{Z_{t}}{Z_{0}} \,.
\label{QJarz}
\end{eqnarray}
 Hence, for  closed systems, the quantum work $W$ characterizes a process where the energy is measured  twice, at time 0 and time $t$  \cite{Talkner1}.
\hfill\break
{\it Remark:}  If we suppose that  $H_{u}$ commutes with $\partial_{u}H_{u}$, then the 
 formula~(\ref{ClosedJara}) can be brought into a simpler form.  From equation~(\ref{Def:WHeisen}), we have
$ W_{u}^{{\mathcal H}}(u) = -\partial_{u}\left(\ln Z_{u}\right)+\beta \left(\partial_{u}H_{u}\right)^{{\mathcal H}}(u)$
  where we explicitly  used that $[H_{u}, \partial_{u}H_{u}] =0$. We then obtain 
  \begin{equation}
  \frac{Z_{t}}{Z_{0}} Tr \left( \pi_{t}A \right)=  Tr\left( \pi_{0}   \, 
 \overrightarrow{\exp }
\left( -\beta \int_{0}^{t}  \left(\partial_{u}H_{u}\right)^{{\mathcal H}}(u)  \, du \right)  A^{{\mathcal H}}(t) \right) \, .
\end{equation}
 The case $A=1$ gives  the H\"anggi-Talkner form \cite{Talkner1}
of the quantum Jarzynski relation for closed systems: 
\begin{equation}
Tr\left( \pi_{0}  \, 
 \overrightarrow{\exp }
\left( -\beta \int_{0}^{t} \left(\partial_{u}H_{u}\right)^{{\mathcal H}}(u)  \, du \right)   \right) 
=\frac{Z_{t}}{Z_{0}} \, .
\end{equation}
Remark that this relation is true if $[H_{u}, \partial_{u}H_{u}] \neq 0$. 
This fact was neglected in \cite{Talkner1} and corrected in an addendum of \cite{CampisiReview}.

\subsection{Steady-State Quantum Fluctuation Dissipation Theorem}

 We now return to the general case of an open system. 
 One main advantage  of the identity~(\ref{Jara})
 is that it implies 
 a  generalized fluctuation-dissipation theorem  valid
 in the vicinity of  a quantum  non-equilibrium steady state.   
 We start with a  Lindbladian  $L_0$, which does not depend on time,   with invariant density-matrix
 given by $\pi_0$. We then   consider  a perturbation of $L_0$  of  the form 
\begin{equation}
  L_{t}= L_0- h^{a}(t) M_{a}  \, .  
 \label{Lpert}
\end{equation}
 The time-dependent
 perturbations  $h^{a}(t)$   are supposed to be small
 and a summation over the repeated index $a$ is understood.
 At first order, the accompanying  density-matrix $\pi_t$, that satisfies 
  $ L_{t}^{\dagger} . \pi_t =0$,  is given by
\begin{equation}
 \pi_t = \pi_0 + h^{a}(t)  \epsilon_a \, , 
 \end{equation}
 where $\epsilon_a$ satisfies
\begin{equation}
  L_{0}^{\dagger} . \epsilon_a =M_{a}^{\dagger} . \pi_0 \, . 
\end{equation}
  The operator  $W_t$, defined in~(\ref{W}),  reads
  \begin{equation}
  W_{t} =  -\dot{h}^{a}(t)  D_{a}  \,\, \hbox{  with } \,\,\, 
   D_{a} =  \pi_0^{-1}  \epsilon_a   \, . 
  \label{def:Da}
  \end{equation} 

 We now take the functional derivative  of the identity~(\ref{Jara})
 w.r.t. $h^a(u)$  for $u<t$. The derivative of the l.h.s. vanishes because
 $\pi_t$  depends only on $h^a(t)$ and not on  $h^a(u)$  for $u<t$. We evaluate
 the derivative of the r.h.s. by using the first order expression~(\ref{def:Da}) for 
 $W_u$. This yields
\begin{equation}
\left. \frac{\delta \left\langle A\left( t \right)\right\rangle  }{\delta h^{a}(u)}
\right| _{h=0}=\frac{d}{du}\left\langle D_{a}(u)A\left( t \right)\right\rangle_{ \pi _{0}}  \, .
\label{TFDG}
\end{equation}
 We emphasize that the expectation value on the r.h.s. is taken
 with respect to the unperturbed density matrix $\pi_0$.
 By choosing  $A_{T}=D_{b}(T)$,
 Eq.~(\ref{TFDG})   becomes structurally similar  to 
 the usual equilibrium fluctuation dissipation theorem.
 This  generalizes  to the quantum case  a  result
 obtained recently  for classical systems
  \cite{Prost,Che} (see  \cite{Weidlich} for  an alternative approach).


\hfill\break
{\it Remark:} In the case of  a closed system  perturbed near equilibrium 
 the  steady-state quantum Fluctuation Dissipation  Theorem~(\ref{TFDG}) 
 reduces to the celebrated Callen-Welton-Kubo relation \cite{Callen,Kubo}.
 The details of the proof  are given in Appendix~\ref{Proof:Callen-Welton}.

\section{Time reversal and a  Quantum Tasaki-Crooks  Relation}
  \label{sec: Tasaki-Crooks}

  Symmetry by   time reversal lies at the heart of many exact identities
  in non-equilibrium  statistical mechanics: indeed, various fluctuation 
 relations can be derived by comparing the averages performed on a given 
 process with those  performed on the time-reversed process. 
 We shall first recall  how  time reversal can be defined for quantum
 Markov processes. Then we shall use this operation to derive a general
 version of the Tasaki-Crooks  Relation, valid for open quantum systems.


\subsection{Time-reversal for  Lindblad dynamics}

 In Quantum Mechanics, time reversal  on the states $\psi$ of the Hilbert space
  is implemented by an anti-unitary operator
 \cite{Schiff} that we denote by $\theta$. The operator $\theta$ is anti-linear
 and it satisfies  $\theta^2=1$,  $ \theta= \theta^{-1}= \theta^{\dagger}$.
 For a spin-0 particle without magnetic field,  $\theta$ can be identified 
 with  the  complex conjugation  operator (i.e.  by  time reversal, the Schr\"odinger
 wave function $\psi$ becomes $\psi^{*}$).
 In presence of  a magnetic field, this  time-inversion
 operation  must be  supplemented by the
 requirement  that the reversed system evolves with  potential vector
  $A^{R}= - A $.   Time reversal  of observables
 (which are operators acting on the  Hilbert space
 of states)
 is implemented by a superoperator
 $K$  that  acts on an operator $X$ as follows \cite{Agarwal,Majewski}:
 \begin{equation}
   K X = \theta X \theta^{-1} \, .
\label{def:actionK}
\end{equation}
 The  superoperator  $K$ is anti-unitary, with  $K^2=1$,  $K=K^{-1}=K^{\dagger}$ 
 and is multiplicative
  i.e.  $K(XY)=K(X)K(Y)$.

 We can now define  time reversal for  a quantum Markov process.   We consider 
 the   case of a  system  with  a constant  Lindbladian $L$ that lies in a 
 steady state with density-matrix $\pi$. Conditions for defining the 
 time-reversed quantum Markov process have been stated by various authors
  \cite{Agarwal,Majewski,Fagnola,Fagnola2}.
  The superoperator $L^{R}$ that governs the
 reversed process is given by 
\begin{equation}
  L^R=   K \pi^{-1} L^{\dagger} \pi K \,,
\label{def:LR}
\end{equation}
 where $\pi^{-1}$  and $\pi$  are  understood   as left-multiplication
 superoperators, {\it i.e.,} if $X$ is an operator we have $ L^R(X)=
 K  (\pi^{-1} L^{\dagger}(\pi (K X)))$.  The condition of {\it
   micro-reversibility} or detailed balance is then  expressed as
 $L^R =L$ which is equivalent to $K \pi^{-1} L^{\dagger} \pi K
 =L$. Note also that this relation and its  conjugate  imply that
 $\pi K=K \pi$, so  $ K(\pi)=\pi$; thus,  the detailed balance
 condition also takes the form $L=\pi^{-1}KL^{\dagger} K\pi$ which is
 identical to the condition given in  \cite{Tal1}
 (relation 4.8).  The  exponentiation of this last formula leads us to the 
 finite time formula  $\pi P_{0}^{T} =
     K\left(P_{0}^{T} \right)^{\dagger} K \pi $ which can be  written as 
\begin{equation}
Tr\left( B^{\dagger}\pi P_{0}^{T}  A \right) = Tr\left(  \left(K
A^{\dagger}\right) \pi  P_{0}^{T} \left( K  B \right) \right) \, 
\label{detbal}
\end{equation} 
for two arbitrary observables $A$ and $B$.  
  Formula~(\ref{detbal}) coincides with
   the definition of a detailed balance
  given by  Majewski  \cite{Majewski}  inspired by Agarwal
 (relation 2.19 of  \cite{Agarwal})  and by   Fagnola et al.  \cite{Fagnola2}.
 Besides, in  \cite{Fagnola2}  a characterization is given of
 the  Lindbladians that satisfy  
  detailed balance. Finally, we  must underline that there exists still another
  definition of quantum detailed balance in the sense of Frigerio et
  al. \cite{Frigerio,Frigerio2} and Alicki \cite{Alicki} which can be
  written in the form
  $L-\pi^{-1}L^{\dagger}\pi=2i\left[H,.\right]$. It is shown  in
  \cite{Fagnola2} that this later  definition is equivalent to the previous ones  when  the
  Hamiltonian and the Kraus operator are even observables, i.e. 
  $K\left(H\right)=H$ and $ K(V_{i})=V_{i}$.
 
  More generally, 
 the fact that  $L^R$ defines {\it  a bona fide} quantum dynamics
 is a non-trivial fact that imposes  stringent conditions on the initial
 Lindbladian $L$ \cite{Fagnola,Fagnola2}. One can readily  verify that the stationary
 density matrix associated with the reversed dynamics is given by 
 $\pi^R =  K   \pi$, because  $(L^R)^{\dagger} (K   \pi) = 0$
 (using the fact that $L 1 = 0$).

 Finally, we consider a non-stationary set-up with a time-dependent process,
 governed by a  Lindbladian
 $L_t$ and study  the process  between the initial time $t=0$
 and a  final time $T$.  We wish to consider  a reversed process that also  runs 
 for times between 0 and $T$.  We emphasize that 
  there is not a unique manner to define  time-inversion, as was already realized
 in the case of classical systems \cite{Chetrite}.  We shall write 
 the time-reversed dynamics  by analogy
 with equation~(\ref{def:LR}) and by the requirement that the 
  accompanying distribution of  the time-reversed system is the 
 time-reversed of the  accompanying distribution   of the original system.
 These  two conditions lead to the following  Lindbladian: 
  \begin{equation}
   L_{t^*}^R=   K \pi_t^{-1} L_{t}^{\dagger} \pi_t K \,\,\,\, 
 \hbox{ with } t^* = T -t \, .
\label{def:LindbReversed}
\end{equation}
  Here again, 
  $\pi_t$ and  $\pi_t^{-1}$ denote  left-multiplication superoperators.
  Using  Eq.~(\ref{def:LindbReversed}), and the relation   $L_{t} 1 = 0$ we find
 $(L_{t^*}^R)^{\dagger} K   \pi_t = 0$.  We thus  obtain   $\pi_{t^*}^R =  K   \pi_t$,
  relating, as desired, the   accompanying distribution of  the time-reversed system
 with that   of the original system.
 By applying Eqs.~(\ref{eq:ExpOrd1}) and (\ref{def:Correlations})
 to the   time-reversed system, we find that  the corresponding   evolution
  superoperator  of the   time-reversed system  is given by  $ P_{s}^{t,R}=
 \overrightarrow{\exp }\left( \int_{s}^{t} du \, L_{u}^R \right)$
 and that  the  multi-time  correlations   are:
\begin{eqnarray} 
\left\langle O_{1}(t_{1})O_{2}(t_{2})\ldots O_{N}(t_{N})\right\rangle^R  
= Tr\left( \pi_{0}^R P_{0}^{t_{1},R} O_{1} P_{t_{1}}^{t_{2},R} O_{2} \ldots 
P_{t_{N-1}}^{t_{N},R} O_{N}\right) \, .
\label{eq:CorrelRenv}
 \end{eqnarray} 
(The superscript $R$ on the l.h.s. recalls  that  correlations are taken
 for the {\it time-reversed} process.) 
 We again emphasize  that the superoperator  $L_{t^*}^R$, given  in equation~(\ref{def:LindbReversed}), 
 must be a well-defined Lindbladian: this a  non-trivial property.  This  property can be ensured  by imposing 
 at each time $t$ the quantum {\it instantaneous} detailed balance condition $L_{t^*}^R = L_{t}$. Here, we
 do not assume detailed balance and we only require the weaker condition that 
 $L_{t^*}^R$  is a  Lindbladian.

\subsection{Proof of the Quantum Tasaki-Crooks Relation}


 Given a scalar $\alpha$, with $0 \le \alpha  \le 1 $, 
 we  introduce two  {\it  $\alpha$-deformed superoperators} 
  $L_{t}(\alpha)$ and $L_{t}^R(\alpha),$ 
 that act  on an observable  $X$  as follows:
\begin{eqnarray}
L_{t}(\alpha)    X &=&   \left( L_{t}  
  + \alpha \pi _{t}^{-1}\partial _{t}\pi _{t} \right) X   \nonumber \\
  \hbox{  and } \,\,\,
L_{t}^R(\alpha)   X &=& \left( L_{t}^R  
  + \alpha (\pi_t^R)^{-1}\partial _{t}\pi_{t}^R \right) X   \,.  
\label{def:LindbDef}
\end{eqnarray}
 The   superoperators   $L_{t}(\alpha)$   interpolate between  
  $L_t$  (the original  Lindbladian) 
and  $L_{t,1}$ [defined in  equation~(\ref{def:Lind1})] when $\alpha$  varies from 0 to 1.
  Similarly,    $L_{t}^R(\alpha)$ is an   interpolation from 
    $L_t^R$ to  $L_{t,1}^R$.

The corresponding   $\alpha$-deformed evolution superoperators are given by 
\begin{equation}
     P_{s}^{t}(\alpha)=
 \overrightarrow{\exp }\left( \int_{s}^{t} du \,  L_{u}(\alpha)\right) 
 \,\,\,\, \hbox{ and  } \,\,\,\, 
  P_{s}^{t,R}(\alpha)=
 \overrightarrow{\exp }\left( \int_{s}^{t} du \,  L_{u}^R(\alpha)\right) \, . 
\label{eq:ExpOrdalpha}
 \end{equation} 

 These modified superoperators  satisfy the following key duality relation, that
 lies at  the heart of the proof of the quantum fluctuation theorem
 (for the  classical analog in which 
  the  dynamics is also  modified with respect to a continuous parameter
  see \cite{JarzPRE,Kurchan2}):
 \begin{equation}
  \pi_0 P_{0}^{T}(\alpha) = 
 \left[    \pi_T  K  P_{0}^{T,R}(1- \alpha) K \right]^{\dagger}  \, .
\label{eq:Conjug1}
\end{equation}
 This relation is proved by the differential equation technique.
 The operator  
$U_t =  \pi_0 P_{0}^{t}(\alpha)  \pi_t^{-1}$ is equal to 1 at $t=0$
 and it  satisfies the following evolution
 equation 
 \begin{equation}
 \partial_t U_t = \pi_0 P_{0}^{t}(\alpha)\pi_t^{-1}  
 \left( \pi_t L_{t}(\alpha) \pi_t^{-1} -  \partial _{t}\pi _{t} \, \pi_t^{-1} \right )
 =  U_t \left( \pi_t L_{t} \pi_t^{-1}  + (\alpha -1)  \partial _{t}\pi _{t} \, \pi_t^{-1}  \right )
 =  U_t \left(K L_{t^*}^R (1- \alpha)  K \right)^{\dagger}  \,,
\end{equation}
where we have used equation~(\ref{def:LindbReversed}) for the last equality.  Hence, 
 we can write
 \begin{equation}
 U_T = \overrightarrow{\exp }\left( \int_{0}^{T} du \,  ( K \, L_{T-u}^R \, K)^{\dagger}
  (1-\alpha)\right)   =   
  \left( K \,  \overrightarrow{\exp }  \int_{0}^{T} dv \,  L_{v}^R 
  (1-\alpha)  \,   K   \right)^{\dagger} =  \left( K  P_{0}^{T,R}(1- \alpha) K \right)^{\dagger} \,,
 \end{equation}
which proves  equation~(\ref{eq:Conjug1}).

 Applying the duality  identity~(\ref{eq:Conjug1})
  to two arbitrary observables 
 $A$ and $B$ and using  the fact that $K$ is multiplicative  and anti-unitary,
 we obtain the relation 
\begin{equation}
Tr\left( B^{\dagger}\pi _{0}P_{0}^{T}(\alpha)  A \right) = Tr\left(
\left(K   A^{\dagger}\right) \pi_{0}^R P_{0}^{T,R}(1-\alpha )  \left( K  B \right) \right) \, . 
\label{fluctuat}
\end{equation}
 This equation  is  the   essence of  the quantum
 fluctuation theorem  and it  expresses a  generalized detailed balance condition. 
  [Indeed, for a system in   a reversible  stationary state i.e.  $\pi_t = \pi$ 
  and  $P_{0}^{T,R}=P_{0}^{T}$,  it 
 becomes identical to the detailed balance condition (\ref{detbal}) 
  \cite{Agarwal,Majewski,Fagnola, Fagnola2}.]  
 Equation~(\ref{fluctuat})   can be brought into the following more familiar form  by
  using the same method as in section~\ref{subsec:ProofJ} 
 (see Appendix~\ref{Proof:QuantFT} for more details):
\begin{eqnarray}
&\left\langle \left( \pi _{0} B\pi _{0}^{-1}\right)^{\dagger}\left( 0 \right)
\overrightarrow{\exp }\left( -\alpha \int_{0}^{T}du \, W_{u}(u)\right)
A(T)\right\rangle  =  \label{QuantFT} \\  & \left\langle \left( \pi _{0}^{R} \left( K A \right)
\left( \pi_{0}^{R}\right) ^{-1}\right)^{\dagger}\left( 0 \right)
\overrightarrow{\exp }\left( -(1- \alpha )\int_{0}^{T}du \, W_{u}^{R}(u)\right)
 \left( K B \right)(T)\right\rangle ^{R}  \, ,  \nonumber 
\end{eqnarray}
where  $W_t^R$ denotes the  reversed 'injected-power' operator,
 given by   $ W_t^R = -(\pi_t^R)^{-1} \partial_t \pi_t^R \,. $
  This   identity  is original and  it implies all the other 
  results described in the present   work. 
  In particular, if we take
 $B = \thickone$ and $\alpha =1$, 
 then  Eq.~(\ref{QuantFT}) is the 
 quantum  Jarzynski-Hatano-Sasa identity Eq.~(\ref{Jara}).
 If we interpret the mean values in  Eq.~(\ref{QuantFT})  as classical
 averages and the operators as commuting c-numbers, then Eq.~(\ref{QuantFT}) becomes
 Crooks'relation.

   \hfill\break
{\it Remark:} For a closed system,  with an  evolution operator $U_{0}^{t}$ given 
  in equation~(\ref{eq:closedU}), we can verify that  the time-reversed
 system~(\ref{def:LindbReversed}) is also closed  with 
  Hamiltonian $H_{t^{*}}^{R}= K  H_{t} $  and  evolution
 operator $U_{0}^{T,R} = K . \left(U_{0}^{T}\right)^{\dagger}$.
 Then,  using  identity~(\ref{eq:Conjug1}) for $\alpha=1$  and 
   the fact that $K$ is multiplicative, we obtain 
 \begin{eqnarray}
P_{0}^{T}(1)A = \pi _{0}^{-1}K\left( P_{0}^{T,R}\right) ^{\dagger }K\pi
_{T} A =\pi _{0}^{-1}KU_{0}^{T,R}K\left( \pi _{T} A \right) \left(
U_{0}^{T,R}\right) ^{\dagger } =   \label{evolfermedef} \\\pi _{0}^{-1}K\left( K\left( \left(
U_{0}^{T}\right) ^{\dagger }\right) K\left( \pi _{T} A \right) K\left(
U_{0}^{T}\right) \right) =\pi _{0}^{-1}\left( U_{0}^{T}\right) ^{\dagger
}\pi _{T} A U_{0}^{T}.
\nonumber
\end{eqnarray}
 Substituting the last  expression   in Eq.~(\ref{fluctuat}) leads to 
\begin{equation}
Tr\left( B^{\dagger}\pi _{0}\pi _{0}^{-1}U_{0}^{T\dagger}
\pi_{T}AU_{0}^{T}\right) = Tr\left( K\left(A^{\dagger }\right)
\pi _{0}^{R}\left(U_{0}^{T,R}\right)^{\dagger}K\left(B\right)U_{0}^{T,R}\right) \, .
  \label{QFTisole}
\end{equation}
 Recalling that 
  $\pi _{0}^{-1}$ and $\pi_{T}$ are given by the 
 Boltzmann law,  the above equation  becomes
 in  the Heisenberg representation denoted by the
 superscript ${\mathcal H}$, 
 \begin{eqnarray}
 Tr\left( B^{\dagger}\pi _{0}\exp (\beta H_{0}^{{\mathcal H} }(0))
\exp \left( -\beta H_{T}^{{\mathcal H} }(T)\right)
 A^{{\mathcal H}}(T)\right)   
  = \frac{Z_{T}}{Z_{0}}
 Tr\left( K\left(A^{\dagger }\right)
\pi _{0}^{R}\left(U_{0}^{T,R}\right)^{\dagger}K\left(B\right)U_{0}^{T,R}\right) \,.
\label{QFTisole2}
\end{eqnarray}
 We emphasize  that for  $B=\thickone$, Eq~(\ref{QFTisole}) is a tautology (because $K$ is anti-unitary), however
 it implies the non-trivial result~(\ref{QFTisole2}): this 
   feature is characteristic of most of the  derivations of the
 work identities. If we  take   $A=B=\thickone$, we retrieve
 the quantum Jarzynski relation for closed systems
 as first found  by Kurchan and Tasaki \cite{Kurchan,Tasaki}.

\section{Conclusion}
\label{sec:Conclusion}

  In this work, we have derived  fluctuations relations for an open
  quantum system   described by a  Lindblad dynamics that  takes into
  account the  interactions  with the  environment as well as
  measurement processes.  We prove  the  fluctuations relations thanks
  to  a suitable deformation of the system's dynamics: this crucial
  technical idea  provides  a truly  unified picture of the
  fluctuations relations, whether classical or quantum,  and does not
  require  to define  the concept of work at the quantum level.
    Quantum Fluctuation Relations for open
    systems are, at present, not  as developed as their  classical
   counterparts. One major difficulty in the quantum realm is the 
    lack of a trajectory picture when coherence and measurements are
    taken into account. To overcome  this
    difficulty,  an {\it unravelling}  of the Lindblad equation
    \cite{Breuer, Gardiner} must be used.  Previous attempts of this idea were
    performed in \cite{Roeck,Roeck2,Dere} and we plan to extend the
    results  of the  present work  by using such unravellings
    \cite{Attal3}.  Another possible extension  of our work would be to  study
   the effect of choosing a  time inversion different from that of
  Eq.~(\ref{def:LindbReversed}): this   may lead to various families of
  fluctuation relations,  as happened  in the classical case
  \cite{Chetrite}.  One could also study 
  particles with non-zero spins such as Dirac spinors.
   Exact  solutions of   specific  models (such as  quantum Brownian
  motion)  may also  provide us with  experimentally testable
  predictions; more precisely,  the formal relation
    between  the classical exclusion process and the Lindblad
    evolution of free fermions in one dimension \cite{Temme, Eisler},
    could allow us  to use, for  open quantum systems, the exact results 
    obtained for  the large deviation functions of 
   various stochastic  processes \cite{Derrida, Mallick}. 
   Finally, the investigation  of
  time-reversal properties of quantum Non-Markovian systems, in which
  the characteristic time scale of the environment can not be
  neglected w.r.t. that of the system,  should  also yield
  interesting  fluctuation relations.

 R. C. thanks  K. Gaw\c{e}dzki for pointing out the fact that 
 the Lindbladian character of Eq.~(\ref{def:LindbReversed}) is non-trivial
 and the relation with detailed balance.
  R. C.  acknowledges the support of the Koshland center for basic research.
 K.M. thanks M. Bauer  and  H. Orland for useful comments
 and  S. Mallick for useful  remarks on the manuscript. 
  Results similar to those presented here
 were also reached  independently by  K. Gaw\c{e}dzki
  and S.  Attal some time ago \cite{Attal}.

\appendix

\section{Proof of  Equation~(\ref{IdentiteFermee})}
 \label{Proof:IdentiteFermee}

  The identity~(\ref{IdentiteFermee})
  is proved by using the differential equation technique.  First, we note  that both sides
 of equation~(\ref{IdentiteFermee}) coincide at $t=0$.  Then, we find that the time derivative of the l.h.s.
 is given by 
$$ \frac{d}{dt} \left\{ \overrightarrow{\exp }
\left( -\int_{0}^{t} W_{u}(u)^{{\mathcal H}}  \, du \right)   \right\} = 
\overrightarrow{\exp }
\left( -\int_{0}^{t} W_{u}(u)^{{\mathcal H}}  \, du \right) ( -  W_{t}(t)^{{\mathcal H}} ) \, . $$
This follows from  the very definition of a time-ordered exponential.

 The time derivative of the r.h.s.
is given by 
 \begin{eqnarray}
 \frac{d}{dt} \left(  (\pi_0 )^{-1}\pi_t^{{\mathcal H}}(t)  \right) &=&  
 (\pi_0 )^{-1} \frac{d}{dt} \left\{  \left( U_{0}^{t}\right)^{\dagger} \,  \pi_t \,  U_{0}^{t} \right\}
 =  (\pi_0 )^{-1}  \left\{ i \left( U_{0}^{t}\right)^{\dagger} \,[H_t,\pi_t  ]   \,  U_{0}^{t}
 +  \left(\frac{d \pi_t}{dt} \right)^{{\mathcal H}}   \right\}   \nonumber \\
   &=&  
  (\pi_0 )^{-1} \pi_t^{{\mathcal H}}(t)
\left\{ \left( \pi_t^{{\mathcal H}}(t) \right)^{-1}\,  \left(\frac{d \pi_t}{dt} \right)^{{\mathcal H}}      \right\}   
=   (\pi_0 )^{-1} \pi_t^{{\mathcal H}}(t) ( -  W_{t}(t)^{{\mathcal H}} )
\end{eqnarray} 
 where we have used the fact that $\pi_t$  commutes with $H_t$. The last equality follows from the
 definition~(\ref{W}) of $W_{t}$. We have thus shown that the  l.h.s. and the r.h.s coincide at $t=0$
 and that they satisfy the same
 first order  differential with respect to time: they are therefore identical for all times.

\section{Proof of the Callen-Welton-Kubo Formula
 using  the steady-state quantum fluctuation dissipation theorem (\ref{TFDG})}
 \label{Proof:Callen-Welton}

 In this Appendix, we  show  that the steady-state
  quantum Fluctuation Dissipation Theorem~(\ref{TFDG}) is equivalent, 
 in the case of  a closed system  perturbed near equilibrium,
 to the  Callen-Welton-Kubo relation \cite{Callen,Kubo}.
  
  We start with a  time-independent Hamiltonian   $H_0$  with invariant density-matrix
 given by $\pi_0 = Z_{0}^{-1}\exp (-\beta H_{0}).$  We then  
  consider  a perturbation of $H_0$  of  the form 
\begin{equation}
  H_{t}= H_0- h^{a}(t) O_{a}   \, .
 \label{H0pert}
\end{equation}
In a closed system, the Lindblad equation reduces to the Liouville
  equation
\begin{equation}
      \partial _{t}\rho _{t} = L_{t}^{\dagger}  \rho _{t} =       - i[H_t,\rho _{t}] \, .
\end{equation} 
 The  accompanying
  density is explicitly  given  by  $\pi _{t}=Z_{t}^{-1}\exp (-\beta H_{t}).$
 Comparing with equation~(\ref{Lpert}), we find  $M_a = i[O_a, .]$.
 
 We now  derive  an  explicit expression for the operators  $D_{a}$ defined
  in equation~(\ref{def:Da}).
 Starting with  the exact formula
\begin{equation}
\exp (-\beta H_{t})= \overrightarrow{\exp} 
\left( h_{t}^{a}\int_{0}^{\beta }d\alpha \exp
(-\alpha H)O_{a}\exp (\alpha H) \right) \, \exp (-\beta H) \, , 
\end{equation}  
  (which can be proved by differentiating both sides w.r.t. $\beta$),  we find at first order \cite{Orland}
\begin{equation}
\exp (-\beta H_{t})=\exp (-\beta H)+h_{t}^{a}\int_{0}^{\beta }d\alpha \exp
(-\alpha H)O_{a}\exp (\alpha H)\exp (-\beta H) +o(h) \, .
\end{equation}
 This  implies that 
\begin{eqnarray}
Z_{t} &=&Tr\left( \exp (-\beta H)\right) +h_{t}^{a}\beta Tr\left(O_{a}\exp
(-\beta H) \right) +o(h)\text{ } \\
&=&Z\left( 1+h_{t}^{a}\beta \left\langle O_{a}\right\rangle _{\pi _{0}}\right) +o(h) \, .
\end{eqnarray}
The first order perturbation of $W_{t}$ is then given by 
 \begin{eqnarray*}
W_{t} &=&\dot{h}_{t}^{a}\left( \beta \left\langle O_{a}\right\rangle_{ \pi _{0}}-\exp
(+\beta H)\int_{0}^{\beta }d\alpha \exp (-\alpha H)O_{a}\exp (\alpha H)\exp
(-\beta H)\right) +o(h)
\end{eqnarray*}
Comparing with equation~(\ref{def:Da}), we  
 obtain  the analytical expression for $D_{a}$: 
\begin{equation}
D_{a}=-\beta \left\langle O_{a}\right\rangle _{ \pi _{0}}
 +\int_{0}^{\beta }d\alpha \exp
(\alpha H)O_{a}\exp (-\alpha H) \, .
\label{ExpreDa}
\end{equation}

 We now transform, using~(\ref{def:Correlations}), the r.h.s. of the 
  quantum Fluctuation Dissipation Theorem~(\ref{TFDG}) as follows:
\begin{eqnarray*}
\frac{d}{du}\left\langle D_{a}(u)A(t)\right\rangle _{ \pi _{0}} &=&\frac{d}{du}%
Tr(\pi _{0}D_{a}P_{u}^{t}A)=-Tr(\pi _{0}D_{a}LP_{u}^{t}A) \\
&=&-(D_{a}^{\dagger}\pi _{0},LP_{u}^{t}A)
=-(L^{\dagger}\left( D_{a}^{\dagger}\pi _{0}\right) ,P_{u}^{t}A) \\
&=&-Tr(\left( L^{\dagger }\left( D_{a}^{\dagger}\pi _{0}\right) \right) ^{\dagger}P_{u}^{t}A)
=-Tr(\pi _{0}\pi _{0}^{-1}\left( L^{\dagger}\left( \pi _{0} D_{a}\right) \right) P_{u}^{t}A) \,.
\end{eqnarray*}
Note that in the first equality, we
use the fact that  $\pi _{0}$ is the invariant density of the unperturbed dynamics. 
  The quantum Fluctuation Dissipation Theorem~(\ref{TFDG}) can thus
 be rewritten as 
  \begin{equation}
\left. \frac{\delta \left\langle A(t)\right\rangle }{\delta h^{a}(u)}%
\right| _{h=0}=\left\langle E_{a}(u)A(t)\right\rangle _{ \pi _{0}}  \label{TFD2}
\end{equation}
where we have defined
\begin{equation}
E_{a}=-\pi _{0}^{-1}L^{\dagger }\left( \pi _{0}D_{a}\right) \,. 
\end{equation}
 From equation~(\ref{ExpreDa}),
 we deduce the  analytical expression of  $E_{a}$: 
\begin{eqnarray*}
E_{a} &=&-\pi _{0}^{-1}L^{\dagger }\left( \pi _{0}D_{a}\right) \\
&=&i\exp (\beta H)\left[ H,\exp (-\beta H)\int_{0}^{\beta }d\alpha \exp
(\alpha H)O_{a}\exp (-\alpha H)\right] \\
&=&i\left[ H,\int_{0}^{\beta }d\alpha \exp (\alpha H)O_{a}\exp (-\alpha H)%
\right] =i\int_{0}^{\beta }d\alpha \exp (\alpha H)[H,O_{a}]\exp (-\alpha H)
\\
&=&i\int_{0}^{\beta }d\alpha \frac{d}{d\alpha }\left( \exp (\alpha
H)O_{a}\exp (-\alpha H)\right) =i\exp (\beta H)O_{a}\exp (-\beta H)-iO_{a} \, .
\end{eqnarray*}
We remark that the terms on the r.h.s. can be interpreted as the analytic continuation
  in imaginary  time (as allowed by the KMS condition \cite{Martin})
 of the  Heisenberg representation with respect to the unperturbed Hamiltonian $H$.
 Thus,  we have
\begin{equation}
E_{a}=iO_{a}^{H}(-i\beta )-iO_{a}^{H}(0) \, .
\end{equation}

Finally, the (\ref{TFD2}) becomes for ($u<T$) 
\begin{eqnarray}
\left. \frac{\delta \left\langle A(t)\right\rangle }{\delta h^{a}(u)}%
\right| _{h=0} &=&iTr\left( \pi _{0}O_{a}^{H}(-i\beta )A^{H}(t-u)\right)
-iTr\left( \pi _{0}O_{a}^{H}(0)A^{H}(t-u)\right) \notag \\
&=&i\left\langle O_{a } (-i\beta)A(t-u)\right\rangle _{\pi _{0}}-i\left\langle
O_{a}(0)A(t-u)\right\rangle _{\pi _{0}}\  \notag \\
&=&i\left\langle O_{a}(0)A(t-u+i\beta )\right\rangle _{\pi _{0}}-i\left\langle
O_{a}(0)A(t-u)\right\rangle _{\pi _{0}}\ \label{TFDreel}
\end{eqnarray}
The last equality follows from the fact that the correlation $\left\langle
X_{s}Y_{t}\right\rangle _{\pi _{0}}$ depends just on
$t-s$.  This equation is the real space version  of the Callen-Welton-Kubo equation.
 The more conventional form is obtained by performing a Fourier Transform with respect
 to time. The susceptibility is defined as (using causality i.e.
$\left. \frac{\delta \left\langle
A(t)\right\rangle }{\delta h^{a}(u)}\right| _{h=0}=0$  if  $u>t$)
\begin{eqnarray}
\chi _{O_{a}A}(w)  
&=&\int_{0}^{\infty }dt\left. \frac{\delta \left\langle A(t)\right\rangle }{%
\delta h^{a}(0)}\right| _{h=0}\exp (iwt)  \notag \\
&=&i(\exp (\beta w)-1)\int_{0}^{\infty }dt\left\langle
O_{a}(0) \, A(t)\right\rangle _{0}\exp (iwt)  \label{DefSucept} \, 
\end{eqnarray}
where we have used  equation~(\ref{TFDreel})
in the last equality. The symmetrized correlation can be written as 
\begin{eqnarray}
C_{O_{a}A}^{S}(w) &=&\frac{1}{2}\int_{-\infty }^{+\infty }dt\exp (iwt) \, Tr(\pi
_{0}O_{a}^{H}(0)A^{H}(t))+\frac{1}{2}\int_{-\infty }^{+\infty }dt\exp
(iwt)Tr(\pi _{0}A^{H}(t)O_{a}^{H}(0))   \notag  \\
&=&\frac{1}{2}\int_{0}^{+\infty }dt \exp (iwt)Tr(\pi
_{0}O_{a}^{H}(0)A^{H}(t))+\frac{1}{2}\int_{0}^{+\infty }dt \exp (-iwt) Tr(\pi
_{0}O_{a}^{H}(0)A^{H}(-t))+  \notag \\
&&\frac{1}{2}\int_{0}^{+\infty }dt\exp (iwt)Tr(\pi _{0}A^{H}(t)O_{a}^{H}(0))+%
\frac{1}{2}\int_{0}^{+\infty }dt \exp (-iwt)Tr(\pi _{0}A^{H}(-t)O_{a}^{H}(0))
 \notag  \\
&=&\frac{1}{2}\int_{0}^{+\infty }dt\exp (iwt)Tr(\pi
_{0}O_{a}^{H}(0)A^{H}(t))+\frac{1}{2}\int_{0}^{+\infty }dt \exp (-iwt)Tr(\pi
_{0}A^{H}(0)O_{a}^{H}(t+i\beta ))+  \notag \\
&&\frac{1}{2}\int_{0}^{+\infty }dt\exp (iwt)Tr(\pi
_{0}O_{a}^{H}(0)A^{H}(t+i\beta ))+\frac{1}{2}\int_{0}^{+\infty }dt
 \exp(-iwt)Tr(\pi _{0}A^{H}(0)O_{a}^{H}(t)) \notag  \\
&=&\frac{1}{2}(1+\exp (\beta w))\int_{0}^{\infty }dt\left\langle
O_{a}(0)A(t)\right\rangle _{0}\exp (iwt)+\frac{1}{2}(1+\exp (-\beta
w))\int_{0}^{\infty }dt\left\langle A(0) O_{a}(t)\right\rangle _{0}\exp (-iwt) \, .
 \,\,\,\,  \,\,\,\, \label{SymmCorr}
\end{eqnarray}
where  we have used a KMS type identity to obtain the 
 third equality:  $Tr(\pi _{0}X^{H}(s)Y^{H}(t))=$ $%
Tr(\pi _{0}Y^{H}(t)X^{H}(s+i\beta ))$.
By using  (\ref{DefSucept}) and  (\ref{SymmCorr}), we obtain 
\begin{eqnarray}
C_{O_{a}A}^{S}(w) &=&\frac{1}{2}(1+\exp (\beta w))\frac{\chi _{O_{a}A}(w)}{%
i(\exp (\beta w)-1)}+\frac{1}{2}(1+\exp (-\beta w))\frac{\chi _{AO_{a}}(-w)}{%
i(\exp (-\beta w)-1)} \\
&=&\frac{\coth (\frac{\beta w}{2})}{2i}\left( \chi _{O_{a}A}(w)-\chi
_{AO_{a}}(-w)\right)  \notag
\end{eqnarray}
Note that in the expression for  $\chi _{AO_{a}}$, the operator $A$ is considered
 to be the perturbation and $O_{a}$  the observable. We thus implicitly suppose
 that $A = A^{\dagger}$.
In particular, if we take  $A =O_{b}$
\begin{equation}
C_{O_{a}O_{b}}^{S}(w)=\frac{\coth (\frac{\beta w}{2})}{2i}
\left( \chi_{O_{a}O_{b}}(w)-\chi _{O_{b}O_{a}}(-w)\right) = \coth (\frac{\beta w}{2}) 
 \operatorname{Im}(\chi_{O_{a}O_{b}}(w)) \,, \label{fdtf}
\end{equation}
 recalling that  $\chi _{AO_{a}}(-w)=\overline{\chi _{AO_{a}}(w)}$.
  For a closed system, an   alternative  proof  showing that  relation (\ref{QJarz2})
  implies Eq.~(\ref{fdtf})
 is given in  \cite{Gaspard1}.   

\section{Derivation of  Equation~(\ref{QuantFT})}
 \label{Proof:QuantFT}

First, we establish 
the following relation, valid for two operators $X$ and $Y$
\begin{equation}
 Tr\left( \pi_{0} Y  P_{0}^{T}(\alpha)  X \right) 
 = \left\langle Y(0)  \,\,  \overrightarrow{\exp }
 \left( -\alpha \int_{0}^{T}du \,  W_{u}(u)\right)  X(T) \right\rangle \,.
\label{def:Dyson}
\end{equation}
 The method to derive this formula 
is identical to the one used in equations~(\ref{DysonSchwinger})
  to (\ref{eq:afterRessum}): we perform  Dyson-Schwinger  expansion  $P_{0}^{T}(\alpha)$
 w.r.t. the deformation parameter $\alpha$,  rewrite the trace as
 a correlation function via 
 Eq.~(\ref{def:Correlations}) and the result is resummed as a time-ordered exponential.
 We recall that  $W_t$ was defined in  equation~(\ref{W}). 
 The same  equation  is true  for the reversed system, with $\alpha$ replaced
 by $1 -\alpha$ and where  the  reversed 'injected-power' operator is given by 
  $ W_t^R = -(\pi_t^R)^{-1} \partial_t \pi_t^R \, $:
\begin{equation}
 Tr\left( \pi_{0} Y  P_{0}^{R,T}(1-\alpha)  X \right) 
 = \left\langle Y(0)  \,\,  \overrightarrow{\exp }
 \left( -(1-\alpha) \int_{0}^{T}du \,  W_{u}^R(u)\right)  X(T) \right\rangle ^{R} \,.
\label{def:Dyson2}
\end{equation}

 Replacing both sides of  equation~(\ref{fluctuat}) by
 the  relations~(\ref{def:Dyson}) and~(\ref{def:Dyson2})  and  
   inserting the
 duality identity~(\ref{eq:Conjug1}) into equation~(\ref{def:Dyson}), leads to 
 the  Fluctuation Relation~(\ref{QuantFT})  for an open quantum  Markovian
 system.

\end{document}